\RequirePackage{fix-cm}
\documentclass[smallextended]{svjour3}       
\smartqed  
\usepackage{graphicx}
\usepackage{amssymb}
\usepackage{epsfig}
\begin{document}

\title{ Logarithmic Corrections in Black Hole Entropy Product Formula 
}
\author{Parthapratim Pradhan}



\institute{    \at
               Department of Physics\\
               Vivekananda Satavarshiki Mahavidyalaya\\
               (Affiliated to Vidyasagar University)\\
               West Midnapur, West Bengal~721513, India \\
               \email{pppadhan77@gmail.com}
          }

\date{Received: date / Accepted: date}

\maketitle

\begin{abstract}
It has been shown by explicit and exact calculation that whenever we have taken the \emph{effects of stable thermal 
fluctuations} the entropy product formula should \emph{not be mass-independent} nor \emph{does it quantized}. 
It has been examined by giving some specific examples for non-rotating and rotating black hole.
\end{abstract}


\section{Introduction}
There has recently been some ongoing intense interest in the thermodynamic product formula both in
the general relativity community \cite{ah09,mv13} and in the String/M-theory community \cite{cgp11,castro12} 
to understand the black hole(BH) entropy \cite{bk72} of multi-horizons at the microscopic level.  Significant
achievements have been made in case of asymptotically flat super-symmetric BHs in four and five
dimension, where the microscopic degrees of freedom could be explained in terms of a
two-dimensional (2D) conformal field theory (CFT)\cite{vafa96}. For microscopic entropy of
extreme rotating BH, the results could be found in \cite{ghss09}.

In case of a regular axi-symmetric and stationary space-time of Einstein-Maxwell gravity
with surrounding matter, the entropy product formula of event horizon (${\cal H}^{+}$)
and Cauchy horizon (${\cal H}^{-}$) is\cite{ah09}
\begin{eqnarray}
{\cal S}_{+} {\cal S}_{-} &=& 4\pi^2J^2+\pi^2Q^4 ~.\label{prKN1}
\end{eqnarray}
In the absence of Maxwell gravity, the product formula reduces to the following form
\begin{eqnarray}
{\cal S}_{+} {\cal S}_{-} &=& 4\pi^2J^2 ~.\label{prK1}
\end{eqnarray}
The interesting point in the above formula is that they all are independent of the 
ADM (Arnowitt-Deser-Misner) mass parameter.

After incorporating the BPS (Bogomol'ni-Prasad-Sommerfeld) states the entropy product formula of 
${\cal H}^{\pm}$ should read
\begin{eqnarray}
{\cal S}_{+} {\cal S}_{-}  &=& 4\pi^2 \left(\sqrt{N_{1}}+\sqrt{N_{2}}\right)\left(\sqrt{N_{1}}-\sqrt{N_{2}}\right)
= N ,\, N\in {\mathbb{N}}, N_{1}\in {\mathbb{N}}, N_{2} \in {\mathbb{N}} \nonumber \\
~.\label{ss}
\end{eqnarray}
where the integers $N_{1}$ and $N_{2}$ are the excitation numbers of the left and
right moving modes of a weakly-coupled 2D CFT. The integers also depend upon exactly all of the BH parameters.
Eq. \ref{ss} indicates that the product of ${\cal H}^{\pm}$ could be expressed in terms of the 
underlying CFT which can be interpreted in terms of a level matching condition. That means the
entropy  product of ${\mathcal H}^{\pm}$ is an integer quantity \cite{fl97} and it also implies that 
it is quantized in nature.

However in this \emph{Letter}, we wish to analyze the general logarithm correction in entropy product formula due to
the effect of thermal fluctuations around equilibrium. Whenever we have taken into account of the quantum 
fluctuations the Bekenstein-Hawking entropy formula  should be corrected that indicates the Bekenstein-Hawking 
entropy product formula of ${\cal H}^{\pm}$  should also be corrected. This is the motivation behind this work. 
We have explicitly examined by giving some specific examples. 

The fact is that certain BHs (in Einstein gravity or other theories of gravity) have larger than the Planck
scale length where the entropy is proportional to the horizon area \cite{bk73,ajak,carlip}. Therefore there 
has been another motivation to find what the leading order corrections in BH entropy as well as in \emph{BH entropy 
product} formula of ${\cal H}^{\pm}$ when one reduces the size of the BH. It has been shown that in case of 
large BHs, the logarithm of the density of states is precisely the BH entropy with the corrections 
$-\frac{3}{2} \ln A$, where $A$ is the area of ${\cal H}^{+}$ \cite{km1,skm}.  Therefore it could be seen that the logarithmic corrections to Bekenstein-Hawking entropy of ${\cal H}^{\pm}$ is a generic property of BHs.

\section{General Logarithmic Corrections to BH Entropy:}
In this section we shall derive the logarithmic corrections to BH entropy (which will be the \emph{master formula} of our investigation) for any thermodynamic system whose \emph{specific heat is 
positive} that means the corresponding canonical ensemble is thermodynamically stable. Thus this calculation is valid for all BHs whose specific heat is being positive.

To determine the leading order corrections to BH entropy of ${\cal H}^{\pm}$ of any thermodynamical
system due to small statistical fluctuations around equilibrium, we should follow the work by Das et al. \cite{psm}.
In the micro canonical thermodynamical system the partition function computed at ${\cal H}^{\pm}$ may 
be defined as
\begin{eqnarray}
 Z_{\pm}(\beta) &=& \int_{0}^{\infty} \rho_{\pm}(E_{\pm}) e^{-\beta_{\pm} E_{\pm}} dE_{\pm}  
~. \label{pf}
\end{eqnarray}
where $T_{\pm}=\frac{1}{\beta_{\pm}}$ is the temperature of the ${\cal H}^{\pm}$. We set Boltzmann constant $k_{B}$ to be unity.

The density of states of ${\cal H}^{\pm}$ can be expressed as an inverse Laplace transformation of
the partition function:
\begin{eqnarray}
 \rho_{\pm}(E_{\pm}) &=& \frac{1}{2 \pi i}\int_{c-i\infty}^{c+i\infty} Z_{\pm}(\beta_{\pm})
 e^{\beta_{\pm} E_{\pm}} d\beta_{\pm} = \frac{1}{2 \pi i}\int_{c-i\infty}^{c+i\infty}
 e^{S_{\pm}(\beta_{\pm})} d{\beta_{\pm}}~. \label{ipf}
\end{eqnarray}
where $c$ is a real constant and we have defined
\begin{eqnarray}
S_{\pm} &=&  \ln Z_{\pm} +\beta_{\pm} E_{\pm}  ~. \label{ents}
\end{eqnarray}
is the exact entropy as a function of temperature, not just its value at equilibrium.

Now close to the equilibrium inverse temperature $\beta_{\pm}=\beta_{0, \pm}$, we may expand the 
entropy function of ${\cal H}^{\pm}$ as
\begin{eqnarray}
S_{\pm}(\beta_{\pm}) &=&  S_{0, \pm}+\frac{1}{2} (\beta_{\pm}-\beta_{0, \pm})^2 S_{0, \pm}'' + ...  ~.\label{sbs0}
\end{eqnarray}
where $S_{0, \pm}: =S_{\pm}(\beta_{0, \pm})$ and $S_{0, \pm}''=\frac{\partial^2 S_{\pm}}{\partial \beta_{\pm}^2}$
at $\beta_{\pm}=\beta_{0, \pm}$.

Substituting (\ref{sbs0}) in (\ref{ipf})
\begin{eqnarray}
\rho_{\pm}(E_{\pm}) &=& \frac{e^{S_{0, \pm}}}{2 \pi i}\int_{c-i\infty}^{c+i\infty}
e^\frac{\left(\beta_{\pm}-\beta_{0, \pm}\right)^2 S_{0, \pm}''}{2} d{\beta_{\pm}}~.\label{rhoe}
\end{eqnarray}

Defining $\beta_{\pm}-\beta_{0, \pm} =i x_{\pm}$ and choosing $c=\beta_{0, \pm}$, $x_{\pm}$ is a real variable and computing a contour integration one can obtain
\begin{eqnarray}
 \rho_{\pm}(E_{\pm}) &=& \frac{e^{S_{0, \pm}}}{\sqrt{2 \pi S_{0, \pm}''}}~.\label{ps0}
\end{eqnarray}
The logarithm of the $\rho_{\pm}(E_{\pm})$ gives the corrected entropy of ${\cal H}^{\pm}$:
\begin{eqnarray}
{\cal S}_{\pm}:  &=&  \ln \rho_{\pm} ={\cal S}_{0, \pm}-\frac{1}{2} \ln S_{0, \pm}''+ ... ~. \label{srho}
\end{eqnarray}

This formula is applicable to all the thermodynamic system considered as a canonical ensemble, including
BH consisting of ${\cal H}^{\pm}$.

Now from the idea of statistical mechanics the mean value of energy \cite{sommer} of any thermodynamical 
system is given by
\begin{eqnarray}
 <E_{\pm}> =-\frac{\partial}{\partial \beta_{\pm}}\ln Z_{\pm} \mid_{\beta_{\pm}=\beta_{0, \pm}} =
 -\frac{1}{Z_{\pm}}\frac{\partial Z_{\pm}}{\partial \beta_{\pm}}\mid_{\beta_{\pm}=\beta_{0, \pm}}
 ~.\label{engpf}
\end{eqnarray}
and the square value of the energy is given by
\begin{eqnarray}
 <E_{\pm}^2> &=& \frac{1}{Z_{\pm}}\frac{\partial^2 Z_{\pm}}{\partial \beta_{\pm}^2}\mid_{\beta_{\pm}=\beta_{0, \pm}}
 ~.\label{eng2pf}
\end{eqnarray}

The fluctuations are equal to the differences between special measured values $E_{\pm}$ and the mean value $<E_{\pm}>$,
or
\begin{eqnarray}
\Delta E_{\pm}=E_{\pm}-<E_{\pm}>  ~.\label{aveng}
\end{eqnarray}
Thus the mean value of the square of the fluctuation in energy is given by
\begin{eqnarray}
(\Delta E_{\pm})^2 &=& <E_{\pm}^2>-<E_{\pm}>^2 \nonumber\\
                   &=&\left[\frac{1}{Z_{\pm}}\frac{\partial^2 Z_{\pm}}{\partial \beta_{\pm}^2}\mid_{\beta_{\pm}=\beta_{0, \pm}}-\frac{1}{Z_{\pm}^2}(\frac{\partial Z_{\pm}}{\partial \beta_{\pm}})^2\mid_{\beta_{\pm}=\beta_{0, \pm}}\right]~.\label{sqeng}
\end{eqnarray}
This may be rewritten as
\begin{eqnarray}
(\Delta E_{\pm})^2 &=& <E_{\pm}^2>-<E_{\pm}>^2 =\frac{\partial^2}{\partial \beta_{\pm}^2}\ln Z_{\pm}
\mid_{\beta_{\pm}=\beta_{0}} =-\frac{\partial <E_{\pm}>}{\partial \beta_{\pm}} \nonumber \\
 &=& T_{\pm}^2 C_{\pm} ~.\label{wsqe}
\end{eqnarray}
The mean square of the fluctuation in energy is seen to be determined by thermodynamic
quantities only. It is proportional to the heat capacity $C_{\pm}$.
This specific heat can be defined as
\begin{eqnarray}
 C_{\pm} & \equiv & \frac{\partial <E_{\pm}>}{\partial T_{\pm}}\mid_{T_{0, \pm}} =
 \frac{1}{T_{\pm}^2}\left[\frac{1}{Z_{\pm}}\frac{\partial^2 Z_{\pm}}{\partial \beta_{\pm}^2}
 \mid_{\beta_{\pm}=\beta_{0, \pm}}-\frac{1}{Z_{\pm}^2}(\frac{\partial Z_{\pm}}{\partial \beta_{\pm}})^2
 \mid_{\beta_{\pm}=\beta_{0, \pm}} \right] \nonumber\\
 &=& \frac{S_{0, \pm}''}{T_{\pm}^2}~.\label{engsh}
\end{eqnarray}
It immediately follows that
\begin{eqnarray}
 S_{0, \pm}'' &=& <E_{\pm}^2>-<E_{\pm}>^2 =C_{\pm}T_{\pm}^2~.\label{ct2}
\end{eqnarray}
Using (\ref{eng2pf}) and (\ref{ct2}) we get the leading order corrections to the black hole entropy is
given by
\begin{eqnarray}
{\cal S}_{\pm}  &=&  \ln \rho_{\pm} ={\cal S}_{0, \pm}-\frac{1}{2} \ln (C_{\pm} T_{\pm}^2)+...  ~. \label{canoetp}
\end{eqnarray}
To compute this entropy of ${\cal H}^{\pm}$, we would replace $T_{\pm}$ by $T_{\pm}$, the Hawking temperature
and then we compute the specific heat $C_{\pm}$ for specific BHs having event horizon (EH) and Cauchy horizon (CH). Without loss of generality, to make definite positivity of specific heat in the 
logarithmic term we have made key assumption that $|C_{\pm}|>0$. Therefore Eq. \ref{canoetp} can 
be re-written as 
\begin{eqnarray}
{\cal S}_{\pm}  &=&  \ln \rho_{\pm} ={\cal S}_{0, \pm}-\frac{1}{2} \ln \left|C_{\pm} T_{\pm}^2\right|+...  ~ \label{ec}
\end{eqnarray}
Thus their product is found to be
$$
{\cal S}_{+} {\cal S}_{-}  =
{\cal S}_{0, +}{\cal S}_{0, -}-\frac{1}{2}\left[{\cal S}_{0, +} \ln \left|C_{-} T_{-}^2\right|+{\cal S}_{0,-}
\ln \left|C_{+} T_{+}^2\right|\right]
$$
\begin{eqnarray}
+\frac{1}{4} \ln \left|C_{+} T_{+}^2\right| \ln \left|C_{-} T_{-}^2\right| +...   \label{prod}
\end{eqnarray}
This is the key result of our investigation.  Now let us see what is the implication of this formula? 
When we  taking consideration the first order correction, it implies that the product is always dependent on mass. Thus the theorem of Ansorg-Hennig\cite{ah09} ``The area (or entropy) 
product formula is independent of mass'' is not universal when we taking into consideration the leading order logarithmic correction. Now we give some specific examples to explain it more detail.

\emph{Example 1}:
First we consider the four dimensional Reissner Nordstr{\o}m (RN) BH. The Bekenstein-Hawking entropy and 
Hawking temperature reads as
\begin{eqnarray}
{\cal S}_{0, \pm} &=& \pi r_{\pm}^2 \\
T_{\pm} &=& \frac{r_{\pm}-r_{\mp}}{4\pi r_{\pm}^2}= \frac{r_{\pm}-r_{\mp}}{4{\cal S}_{0, \pm}}  ~.\label{rnHaw}
\end{eqnarray}
where $r_{\pm}=M \pm \sqrt{M^2-Q^2}$, $M$ and $Q$ are mass and charge of the BH respectively.

The specific heat is given by
\begin{eqnarray}
 C_{\pm} &=& \frac{\partial M}{\partial T_{\pm}}=2\pi r_{\pm}^2 \frac{r_{\pm}-r_{\mp}}{3r_{\mp}-r_{\pm}}=
 2\frac{r_{\pm}-r_{\mp}}{3r_{\mp}-r_{\pm}}{\cal S}_{0, \pm} ~. \label{spRN1}
\end{eqnarray}
Here, $|C_{\pm}|>0$ when $\left|\frac{r_{\pm}-r_{\mp}}{3r_{\mp}-r_{\pm}}\right|>0$ 
\footnote{It should be noted that using properties of 
symmetry in nature of $r_{\pm}$, one can find the following thermodynamic quantities at  ${\cal H}^{-}$:
${\cal A}_{-} = {\cal A}_{+}|_{r_{+}\leftrightarrow r_{-}},\,\, {\cal S}_{-}={\cal S}_{+}|_{r_{+}\leftrightarrow r_{-}},\,\, 
T_{-} = -T_{+}|_{r_{+}\leftrightarrow r_{-}}\,\, and \,\, C_{-} = C_{+}|_{r_{+}\leftrightarrow r_{-}}$}.

Therefore, the entropy correction is given by
\begin{eqnarray}
{\cal S}_{\pm}  &=&  \ln \rho_{\pm} ={\cal S}_{0, \pm}+\frac{1}{2} \ln {\cal S}_{0, \pm}-\frac{1}{2}
\ln \left|\frac{(r_{\pm}-r_{\mp})^3}{8(3r_{\mp}-r_{\pm})}\right|+...  ~ \label{entRn}
\end{eqnarray}
It could be seen that the leading order corrections are logarithmic. Their product is found to be
$$
{\cal S}_{+} {\cal S}_{-}={\cal S}_{0, +}{\cal S}_{0, -}
+\frac{1}{2}\left[{\cal S}_{0, +} \ln {\cal S}_{0,-} +{\cal S}_{0,-} \ln {\cal S}_{0,+}\right]
+\frac{1}{4} \ln {\cal S}_{0,+} \ln {\cal S}_{0,-} \nonumber
$$
$$
-\frac{1}{2}\left[{\cal S}_{0, +}\ln \left|\frac{(r_{-}-r_{+})^3}{8(3r_{+}-r_{-})}\right| +
{\cal S}_{0, -}\ln \left|\frac{(r_{+}-r_{-})^3}{8(3r_{-}-r_{+})}\right| \right]
$$
$$
-\frac{1}{4}\left[{\cal S}_{0, +}\ln \left|\frac{(r_{-}-r_{+})^3}{8(3r_{+}-r_{-})}\right| +
{\cal S}_{0, -}\ln \left|\frac{(r_{+}-r_{-})^3}{8(3r_{-}-r_{+})}\right|\right]
$$
\begin{eqnarray}
+\frac{1}{4} \ln \left|\frac{(r_{+}-r_{-})^3}{8(3r_{-}-r_{+})}\right|\ln \left|\frac{(r_{-}-r_{+})^3}{8(3r_{+}-r_{-})}\right|
+ ...   \label{pd1}
\end{eqnarray}
It should be mention that the first term is mass independent but when first order correction is taking into account the product is mass dependent for RN BH. 

Now let us take another example say Kehagias-Sfetsos(KS) BH \cite{ks09} in Ho\v{r}ava-Lifshitz gravity
\cite{ph9a,ph9b,ph9c,ym09,lmp09,mk10,cc9a,cc9b}.

\emph{Example 2}:
The  entropy,  Hawking temperature and specific heat  for KS BH\cite{horava15} are
\begin{eqnarray}
{\cal S}_{0, \pm} &=& \pi r_{\pm}^2 = \pi \left[ 2M^2 -\frac{1}{2\omega} \pm 2M\sqrt{M^2-\frac{1}{2\omega}} \right],
\end{eqnarray}
where $\omega$ is coupling constant, $r_{+}$ is called EH and $r_{-}$ is called CH.
\begin{eqnarray}
T_{\pm} &=& \frac{\omega (r_{\pm}- r_{\mp})}{4\pi(1+\omega r_{\pm}^2)}, ~\label{M9}
\end{eqnarray}
and
\begin{eqnarray}
C_{\pm} &=& \frac{2 \pi}{\omega} \frac{(2\omega r_{\pm}^2-1)\left(1+\omega r_{\pm}^2\right)^2}
{1+5\omega r_{\pm}^2-2\omega^2 r_{\pm}^4}  .~\label{c5}
\end{eqnarray}
Here, $|C_{\pm}|>0$ when $r_{\pm}^2>\frac{1}{2\omega}$ and $|1+5\omega r_{\pm}^2-2\omega^2 r_{\pm}^4|>0$.

Therefore, the entropy correction for KS BH  is given by
\begin{eqnarray}
{\cal S}_{\pm}  &=&  \ln \rho_{\pm} ={\cal S}_{0, \pm}-\frac{1}{2} \ln \left| \frac{\omega}{4\pi} 
\frac{(2\omega r_{\pm}^2-1)(r_{\pm}-r_{\mp})^2}{1+5\omega r_{\pm}^2-2\omega^2 r_{\pm}^4 }\right|+...  ~ \label{c6}
\end{eqnarray}

and their product yields
$$
{\cal S}_{+} {\cal S}_{-}={\cal S}_{0, +}{\cal S}_{0, -} \nonumber
$$
$$
-\frac{1}{2}\left[{\cal S}_{0, +}\ln \left | \frac{\omega}{4\pi}  \frac{(2\omega r_{-}^2-1)(r_{-}-r_{+})^2}
{1+5\omega r_{-}^2-2\omega^2 r_{-}^4} \right | +
{\cal S}_{0, -}\ln \left | \frac{\omega}{4\pi}  \frac{(2\omega r_{+}^2-1)(r_{+}-r_{-})^2}{1+5\omega r_{+}^2-2\omega^2 r_{+}^4 } 
\right| \right] $$
\begin{eqnarray}
+\frac{1}{4} \ln\left|\frac{\omega}{4\pi}\frac{(2\omega r_{+}^2-1)(r_{+}-r_{-})^2}{1+5\omega r_{+}^2-2\omega^2 r_{+}^4}  
\right| \ln \left|\frac{\omega}{4\pi}  \frac{(2\omega r_{-}^2-1)(r_{-}-r_{+})^2}{1+5\omega r_{-}^2-2\omega^2 r_{-}^4}\right|
+ ...   \label{pd2}
\end{eqnarray}
Again it indicates that when the logarithmic correction is taken into account the entropy product is not universal. 


Now we are moving to the AdS space, see what happens there? First we consider Schwarzschild-AdS space-time.

\emph{Example 3}:
The only one physical horizon \cite{mv13} is given by 
\begin{eqnarray}
{r}_{ +} &=&\frac{2\ell}{\sqrt{3}} sinh \left[\frac{1}{3} sinh^{-1}\left(3\sqrt{3}\frac{M}{\ell}\right)\right] ~\label{ehd}
\end{eqnarray}
Therefore the entropy of ${\cal H}^{+}$ is 
\begin{eqnarray}
 &=& \pi \left(\frac{2\ell}{\sqrt{3}} 
sinh\left[\frac{1}{3} sinh^{-1}\left(3\sqrt{3}\frac{M}{\ell}\right)\right]\right)^2 ~\label{pd3}
\end{eqnarray}
where $\Lambda=-\frac{3}{\ell^2}$ is cosmological constant.
The BH temperature is given by 
\begin{eqnarray}
T_{+} &=& \frac{1}{4\pi r_{+}} \left(1+ 3\frac{r_{+}^2}{\ell^2} \right) ~\label{pd4}
\end{eqnarray}
and the specific heat is given by 
\begin{eqnarray}
C_{+} &=& 2 \pi r_{+}^2 \left[\frac{3\frac{r_{+}^2}{\ell^2}+1}{3\frac{r_{+}^2}{\ell^2}-1} \right].~\label{c7}
\end{eqnarray}
Here, $|C_{+}|>0$ when $r_{+}>\left|\frac{\ell}{\sqrt{3}}\right|$. Since in this case only one physical horizon that is EH, therefore the entropy correction formula \footnote{In the limit $\ell \rightarrow \infty$, one obtains the Schwarzschild BH. In this case the horizon is at 
$r_{+}=2M$ and the specific heat is negative i.e. $C_{+}=-2\pi r_{+}^2$, to make it positive definite we have taken the 
$|C_{+}|=8\pi M^2$ to avoid the divergence in the logarithm term. Therefore the log correction should be 
${\cal S}_{+}=4\pi M^2+\frac{1}{2} \ln|8\pi|$. This also implies that the correction should mass dependent. Thus it is not 
universal.} should be 
\begin{eqnarray}
{\cal S}_{+}  &=& {\cal S}_{0, +}-\frac{1}{2} \ln \left|\frac{\left(3\frac{r_{+}^2}{\ell^2}+1 \right)^3}
{8\pi\left(3\frac{r_{+}^2}{\ell^2}-1 \right)} \right|+...  
~. \label{sads}
\end{eqnarray}
Although it is isolated case but it has still both log correction term and without log correction term depends upon the mass parameter so it is not quantized as well as not universal.

Now we consider the RN-AdS case\cite{mv13}. 

\emph{Example 4}:
The quartic Killing horizon equation becomes
\begin{eqnarray}
 r^4 +\ell^2r^2-2M\ell^2r+Q^2\ell^2 &=& 0 ~.\label{rn1}
\end{eqnarray}
There are at least two physical horizons namely EH, $r_{+}$ and CH, $r_{-}$. The entropy is given by 
\begin{eqnarray}
{\cal S}_{0, \pm} &=& \pi r_{\pm}^2 ~.\label{eq19}
\end{eqnarray}
The BH temperature of ${\cal H}^{\pm}$ should read 
\begin{eqnarray}
T_{\pm} &=& \frac{1}{4\pi r_{\pm}} \left( 3\frac{r_{\pm}^2}{\ell^2}-\frac{Q^2}{r_{\pm}^2}+1\right) ~\label{rn2}
\end{eqnarray}
and the specific heat is given by
\begin{eqnarray}
C_{\pm} &=& 2 \pi r_{\pm}^2 \left[\frac{3\frac{r_{\pm}^2}{\ell^2}-\frac{Q^2}{r_{\pm}^2}+1}
{3\frac{r_{\pm}^2}{\ell^2}+3\frac{Q^2}{r_{\pm}^2}-1} \right].~\label{rn3}
\end{eqnarray}
The specific heat $|C_{\pm}|>0$ definite when $\left|\frac{3\frac{r_{\pm}^2}{\ell^2}-\frac{Q^2}{r_{\pm}^2}+1}
{3\frac{r_{\pm}^2}{\ell^2}+3\frac{Q^2}{r_{\pm}^2}-1} \right|>0$. 

Therefore the logarithmic correction turns out to be 
\begin{eqnarray}
{\cal S}_{\pm}  &=& {\cal S}_{0, \pm}-\frac{1}{2} \ln \left|\frac{\left(3\frac{r_{+}^2}{\ell^2}-\frac{Q^2}{r_{+}^2}+1 \right)^3}
{8\pi\left(3\frac{r_{+}^2}{\ell^2}+3\frac{Q^2}{r_{+}^2}-1 \right)} \right|+...  
~. \label{rn4}
\end{eqnarray}
So we could not write the explicit product formula but it follows that the product of log correction should be mass dependent.

Finally, we consider the simple rotating Kerr BH:

\emph{Example 5}:

The  BH entropy and BH temperature are 
\begin{eqnarray}
{\cal S}_{0, \pm} &=& \pi \left(r_{\pm}^2+a^2\right) \\
T_{\pm} &=& \frac{r_{\pm}-r_{\mp}}{4\pi \left(r_{\pm}^2+a^2\right)}= \frac{r_{\pm}-r_{\mp}}{4{\cal S}_{0, \pm}}  
~.\label{krt}
\end{eqnarray}
where $r_{\pm}=M \pm \sqrt{M^2-a^2}$, $M$ and $a=\frac{J}{M}$ are mass and spin parameter of the BH respectively.

The specific heat \cite{davies} is given by
\begin{eqnarray}
 C_{\pm} &=& \frac{8MT_{\pm}{\cal S}_{\pm}^3}{J^2-8T_{\pm}^2{\cal S}_{\pm}^3} ~. \label{ckr}
\end{eqnarray}
Here, $|C_{\pm}|>0$ when $\left|\frac{8MT_{\pm}{\cal S}_{\pm}^3}{J^2-8T_{\pm}^2{\cal S}_{\pm}^3}\right|>0$.

Finally the logarithmic correction is found to be 
\begin{eqnarray}
{\cal S}_{\pm}  &=& {\cal S}_{0, \pm}-\frac{1}{2} \ln \left|\frac{8MT_{\pm}^3{\cal S}_{\pm}^3}
{J^2-8T_{\pm}^2{\cal S}_{\pm}^3}\right|+...  ~. \label{kr4}
\end{eqnarray}
Once again when the logarithmic correction is taken into account the entropy product of ${\cal H}^{\pm}$ 
is not mass-independent. 

Lastly, we consider the Kerr-Newman BH:

\emph{Example 6}:

The  BH entropy and BH temperature are 
\begin{eqnarray}
{\cal S}_{0, \pm} &=& \pi \left(r_{\pm}^2+a^2\right) \\
T_{\pm} &=& \frac{r_{\pm}-r_{\mp}}{4\pi \left(r_{\pm}^2+a^2\right)}= \frac{r_{\pm}-r_{\mp}}{4{\cal S}_{0, \pm}}  
~.\label{kn1}
\end{eqnarray}
where $r_{\pm}=M \pm \sqrt{M^2-a^2-Q^2}$, $M$, $a$ and $Q$ are mass, spin parameter and charge of the BH respectively.

The specific heat \cite{davies} is 
\begin{eqnarray}
 C_{\pm} &=& \frac{8MT_{\pm}{\cal S}_{\pm}^3}{J^2+\frac{Q^2}{4}-8T_{\pm}^2{\cal S}_{\pm}^3} ~. \label{ckn}
\end{eqnarray}
Here, $|C_{\pm}|>0$ when $\left|\frac{8MT_{\pm}{\cal S}_{\pm}^3}{J^2+\frac{Q^2}{4}-8T_{\pm}^2{\cal S}_{\pm}^3}\right|>0$.

Finally the logarithmic correction is found to be 
\begin{eqnarray}
{\cal S}_{\pm}  &=& {\cal S}_{0, \pm}-\frac{1}{2} \ln \left|\frac{8MT_{\pm}^3{\cal S}_{\pm}^3}
{J^2+\frac{Q^2}{4}-8T_{\pm}^2{\cal S}_{\pm}^3}\right|+...  ~. \label{kn4}
\end{eqnarray}
Once once again when the logarithmic correction is taken into account the entropy product of ${\cal H}^{\pm}$ for 
Kerr-Newman BH \cite{ah09,pp14} is not mass-independent. 

Thus we can conclude from the above examples that when the logarithmic correction is considered the area (or entropy) product formula should \emph{not be universal}. This result should be valid for spherically symmetric regular\cite{grg16} BH, and KN-AdS BH \cite{jh} also.

\section{\label{dis} Conclusion:}
In this work, we have calculated the general logarithmic corrections in black hole entropy product formula by considering the stable thermal fluctuations around the equilibrium. We examined by giving  some specific examples that whenever we taking the first order correction the entropy product formula should not be quantized nor does it mass-independent. We have made a key assumptions to avoid the divergence in the logarithm term is that the absolute value of specific heat for all horizons is always positive. All the derivations in this letter is valid for non-rotating BHs as well as rotating BHs. For extreme BH, $T_{\pm}=0$, hence the thermal fluctuation analysis ceases to be valid only for large quantum fluctuations.

\bibliographystyle{model1-num-names}


\end{document}